\documentstyle[12pt]{article}
\begin{document}
\title{
\bf Semirelativistic quark-antiquark potential, mass spectra and
weak decays into excited states of heavy-light mesons}
\author{ D. Ebert
\\
\small\it Institut f\"ur Physik, Humboldt--Universit\"at zu Berlin,\\
\small\it Invalidenstr.110, D-10115 Berlin, Germany\\
\\
R. N. Faustov and V. O. Galkin \\
\small\it Russian Academy of Sciences, Scientific Council for
Cybernetics,\\
\small\it Vavilov Street 40, Moscow 117333, Russia}
\date{}
\maketitle
\begin{abstract}
Relativistic properties of $q\bar q$ potentail, mass spectra of
orbitally and radially excited $B$ and $D$ mesons as well as
semileptonic decays of $B$ mesons to orbitally excited $D$ mesons
are discussed in the framework of the relativistic quark model
based on the quasipotential approach.
\end{abstract}

{\bf 1. Relativistic properties of $q\bar q$ potential.}
In preceding papers \cite{gmf,fg} we have developed the relativistic
quark model with the ($q\bar q$) potential consisting of the
perturbative one-gluon exchange part and a nonperturbative one which
is a mixture of the Lorentz scalar and vector confining potentials:
\begin{eqnarray}
\label{qpot}
V({\bf p,q};M)&=&\bar{u}_a({\bf p})
\bar{u}_b({\bf-p})\Bigg\{\frac{4}{3}\alpha_sD_{ \mu\nu}({
k})\gamma_a^{\mu}\gamma_b^{\nu}\cr
& & +V^V_{\rm conf}({k})\Gamma_a^{\mu}
\Gamma_{b;\mu}+V^S_{\rm conf}({
k})\Bigg\}u_a({\bf q})u_b({\bf-q}),
\end{eqnarray}
where $k=p-q$,  $D_{\mu\nu}$ is the gluon propagator in the Coulomb gauge
and $\Gamma_{\mu}$ is the effective vector long-range vertex, containing
both the Dirac and Pauli terms
\begin{equation}
\label{vert}
\Gamma_{\mu}=\gamma_{\mu}+
\frac{i\kappa}{2m}\sigma_{\mu\nu}k^{\nu},
\end{equation}
$u_{a,b}({\bf p})$ are the Dirac bispinors. 
The parameter $(1+\kappa)$ can be treated as the nonperturbative
(long-range) chromomagnetic moment of the quark and $\kappa$ as its
anomalous part (flavour independent).

In the nonrelativistic limit the Fourier transform of eq.~(\ref{qpot})
gives the static potential
\begin{equation}
\label{stat}
V_0(r)= V_{\rm Coul}(r)+V^S_{\rm conf}(r)+V^V_{\rm conf}(r),
\end{equation}
where
$$V_{\rm Coul}(r)=-\frac43\frac{\alpha_s}{r},$$
\begin{eqnarray}
\label{lin}
&&V^V_{\rm conf}(r)=(1-\varepsilon)(Ar+B);\qquad
V^S_{\rm conf}(r)=\varepsilon (Ar +B),
\end{eqnarray}
here $\varepsilon$ is the mixing parameter.

Now assuming that both quarks are heavy enough we evaluate the $(v^2/c^2)$
relativistic corrections to the static potential (\ref{stat}), (\ref{lin})
paying special attention to retardation effects. 
The Fourier transform of the linear potential $Ar$ in
the momentum space looks like:
\begin{equation}
\label{four}
A\int {\rm d}^3 r re^{-i{\bf k\cdot r}}=-A\frac{8\pi}{|{\bf k}|^4},
\quad {\bf k}={\bf p}-{\bf q}.
\end{equation}
The natural (though not unique) relativistic extension (dependent
only on the four-momentum transfer) of expression (\ref{four}) is to
substitute $(-{\bf k}^2)\to (k_0^2-{\bf k}^2)$. 
Now  we should choose the procedure of fixing $k_0$.
On the mass shell due to energy conservation we have  $k_0=0$. So $k_0$
may be considered as the measure of deviation either from  the mass shell
or from the energy shell. We choose the second possibility and set  $k_0$
equal to $\epsilon_a({\bf p}) -\epsilon_a({\bf q})$ or to $\epsilon_b({\bf
q})-\epsilon_b({\bf p})$. Then in the symmetrized form  we get
\begin{eqnarray} \label{ko}
&&k_0^2=-(\epsilon_a({\bf p})-\epsilon_a({\bf q}))(\epsilon_b({\bf p})
-\epsilon_b({\bf q})),\qquad
\epsilon_{a,b}({\bf p})=\sqrt{{\bf p}^2+m_{a,b}^2}\nonumber.
\end{eqnarray}
This form is not unique and other possible expressions for $k_0^2$ are
discussed in \cite{g,om}. In favour of choice (\ref{ko}) we mention the
following arguments. It is well-known \cite{ch} that for the one-photon
exchange contribution in QED only choice (\ref{ko}) in the Feynman
(diagonal) gauge leads to the same correct result (the Breit-Fermi
Hamiltonian) as the prescription $k_0=0$ in the Coulomb (or transverse
Landau) gauge. The same is naturally true for the one-gluon exchange
contribution in QCD. Moreover as shown in ref.~\cite{ch} for any
effective vector potential generated by a vector exchange and its
couplings to conserved vector currents (vertices) there is the so-called
instantaneous gauge which plays the role of the Coulomb gauge. In the
instantaneous gauge the prescription $k_0=0$ reproduces the same result
as the expansion in $k_0^2$ fixed by eq.~(\ref{ko}) in the diagonal
gauge used here. The other reason to utilize prescription (\ref{ko})
is the reproduction of the correct Dirac limit in this case \cite{om}.

Carring out $p^2/m^2$ epansion of the $(q\bar q)$ potential (\ref{qpot})
in configuration space we obtain for the spin-independet part the following
expression \cite{gmf,pot}:
\begin{equation}
\label{spind}
V_{\rm SI}(r)=V_0(r) + V_{\rm VD}(r)+\frac18\left(\frac{1}{m_a^2}+
\frac{1}{m_b^2}\right)
\Delta\big[V_{\rm Coul}(r) 
 +(1+2\kappa)V^V_{\rm
conf}(r)\big],
\end{equation}
where $V_0(r)$ is given by eqs.~(\ref{stat}), (\ref{lin}).
The velocity-dependent part $V_{\rm VD}(r)$ can be presented in the
form
\begin{eqnarray}
\label{vdrel}
V_{\rm VD}(r)&=&\frac{1}{m_am_b}\left\{{\bf p}^2V_{bc}(r)+\frac{({\bf p
\cdot r})^2}{r^2}V_c(r)\right\}_W\cr 
&&+\left(\frac{1}{m_a^2}+\frac{1}{m_b^2}\right)\left\{{\bf p}^2 V_{de}
(r) -\frac{({\bf p\cdot r})^2}{r^2}V_e(r)\right\}_W
\end{eqnarray}
with
\begin{eqnarray}
\label{coef}
&&V_{bc}(r)=-\frac{2\alpha_s}{3r}+\left(\frac12-\varepsilon\right)Ar
+(1-\varepsilon)B;\qquad
V_c(r)=-\frac{2\alpha_s}{3r}-\frac12Ar;\cr
&&V_{de}(r)=-\frac{\varepsilon}{2}Ar+\left(\frac14-\frac{\varepsilon}{2}
\right) B;\qquad\quad V_e(r)=0.
\end{eqnarray}
Now we are able to test the fulfilment of the exact Barchielli, Brambilla,
Prosperi (BBP)
relations \cite{BBP}, which follow from the Lorentz invariance
of the Wilson loop. In our notations these relations look like
\begin{equation}
\label{re}
V_{de}-\frac12 V_{bc}+\frac14V_0=0; \quad
V_e+\frac12 V_c+\frac{r}{4}\frac{{\rm d} V_0}{{\rm d} r}=0
\end{equation}
One can easily find that the functions (\ref{coef}) identically
satisfy relations (\ref{re}) independently of values of the parameters
$\varepsilon$ and $\kappa$. This is a highly nontrivial result. For
the perturbative one-gluon-exchange part of $V_{\rm VD}$ our expressions
for $V_b$, $\dots$, $V_e$ are the same as in \cite{BBP}, but for
the confining (long-range) part they are different. 
The terms with the Laplacian in (\ref{spind})
coincide only for $\kappa=0$ and $\varepsilon=0$, i.~e. for
purely vector confining interaction without the Pauli term in the vertex
(\ref{vert}). Our expressions (\ref{spind})  for purely
vector ($\varepsilon=0$) and purely scalar ($\varepsilon=1$)
interactions and for $\kappa=0$ coincide with those of ref.~\cite{om}
except for the constant $B$ term. Our $B$ term for $\varepsilon=1$
(scalar potential) is the same as in \cite{BBP}. The $B$ term from
ref.~\cite{om} does not satisfy the BBP relations (it gives
contribution $-B/2$ only to $V_{de}$). Our
result  for the scalar ($\varepsilon=1$) confining potential
also differs from the one obtained in ref.~\cite{bg}, where the
prescription $k_0=0$ was used and as a result the contribution of
retardation was lost. The differences between our results and the
results presented in ref.~\cite{bv} originate from the use of specific
models such as minimal area law, flux tube, dual superconductivity
and stochastic vacuum.

The spin-dependent part of our potential \cite{gmf,pot} (for $\kappa=-1$)
completely coincides with the one
found in refs.~\cite{b,BBP}. The Gromes relation is identically
fulfilled. Our result supports the conjecture that the long-range
confining forces are dominated by chromoelectric interaction and
that the chromomagnetic interaction vanishes. It is also in
accord with the flux tube and dual superconductivity picture \cite{bv,b}. 

{\bf 2. Mass spectra of heavy-light mesons.}
Many different approaches have been used for the calculation of
orbital and radial excitations of heavy-light mesons \cite{gi,ehq,i}. 
However,
almost in all of them the expansion in inverse powers not only of
the heavy quark mass ($m_Q$) but also  in inverse powers of the 
light quark mass  ($m_q$) is carried out. 
The estimates of the light quark velocity in these mesons show that
the light quark is highly relativistic ($v/c\sim 0.7\div 0.8$). Thus
the nonrelativistic approximation is not adequate for the light quark 
and one cannot guarantee the numerical accuracy of the expansion in inverse
powers of the light quark mass. Here we present the results of the
calculation of the masses of orbitally and radially excited $B$ and
$D$ mesons without employing the expansion in  $1/m_q$
(see \cite{exc} for details). Thus the light
quark is treated fully relativistically. Concerning the heavy quark 
we apply the  expansion in $1/m_Q$ up to the first order. Our
numerical results are presented in Tables~\ref{md}-\ref{mbs}.

\begin{table}[hbt]
\caption{Mass spectrum of $D$ mesons with the account of $1/m_Q$
corrections in comparison with other quark model predictions and
experimental data. All masses are given in GeV. We use the notation
$(nL_J)$ for meson states, where $J$ is the total angular momentum
of the meson.}
\label{md}

\begin{tabular}{ccccc}
\hline
State& our & \cite{gi} & \cite{i}&  experiment \cite{PDG,orb}\\
\hline
$1S_0$ & 1.875 & 1.88 &       & 1.8645(5) \\
$1S_1$ & 2.009 & 2.04 &       & 2.0067(5) \\
$1P_2$ & 2.459 & 2.50 & 2.460 & 2.4589(20) \\
$1P_1$ & 2.414 & 2.47 & 2.415 & 2.4222(18) \\
$1P_1$ & 2.501 & 2.46 & 2.585 &       \\
$1P_0$ & 2.438 & 2.40 & 2.565 &       \\
$2S_0$ & 2.579 & 2.58 &       &       \\
$2S_1$ & 2.629 & 2.64 &       & 2.637(9) ? \\
\hline
\end{tabular}
\end{table}

\begin{table}[hbt]
\caption{Mass spectrum of $D_s$ mesons with the account of $1/m_Q$
corrections in comparison with other quark model predictions and
experimental data. All masses are given in GeV. }
\label{mds}

\begin{tabular}{ccccc}
\hline
State& our & \cite{gi} & \cite{ehq} & experiment \cite{PDG,orb} \\
\hline
$1S_0$ & 1.981 & 1.98   &       & 1.9685(6)\\
$1S_1$ & 2.111 & 2.13   &       & 2.1124(7)\\
$1P_2$ & 2.560 & 2.59   & 2.561 & 2.5735(17) \\
$1P_1$ & 2.515 & 2.56   & 2.526 & 2.535(4) \\
$1P_1$ & 2.569 & 2.55   &       &       \\
$1P_0$ & 2.508 & 2.48   &       &       \\
$2S_0$ & 2.670 & 2.67   &       &       \\
$2S_1$ & 2.716 & 2.73   &       &       \\
\hline
\end{tabular}
\end{table}

\begin{table}[hbt]
\caption{Mass spectrum of $B$ mesons with the account of $1/m_Q$
corrections in comparison with other quark model predictions and
experimental data. All masses are given in GeV. }
\label{mb}

\begin{tabular}{cccccc}
\hline
State & our & \cite{gi} & \cite{i} & \cite{ehq} & experiment
\cite{PDG,orb}\\
\hline
$1S_0$ & 5.285 & 5.31 &       &      & 5.2792(18)\\
$1S_1$ & 5.324 & 5.37 &       &      & 5.3248(18) \\
$1P_2$ & 5.733 & 5.80 & 5.715 & 5.771& 5.730(9) \\
$1P_1$ & 5.719 & 5.78 & 5.700 & 5.759&        \\
$1P_1$ & 5.757 & 5.78 & 5.875 &      &      \\
$1P_0$ & 5.738 & 5.76 & 5.870 &      &        \\
$2S_0$ & 5.883 & 5.90 &       &      &       \\
$2S_1$ & 5.898 & 5.93 &       &      & 5.90(2) ? \\
\hline
\end{tabular}
\end{table}
 
\begin{table}[hbt]
\caption{Mass spectrum of $B_s$ mesons with the account of $1/m_Q$
corrections in comparison with other quark model predictions and
experimental data. All masses are given in GeV. }
\label{mbs}

\begin{tabular}{ccccc}
\hline
State & our & \cite{gi} & \cite{ehq} & experiment \cite{PDG,orb}\\
\hline
$1S_0$ & 5.375 & 5.39 &       & 5.3693(20)\\
$1S_1$ & 5.412 & 5.45 &       & 5.416(4) ? \\
$1P_2$ & 5.844 & 5.88 & 5.861 & 5.853(15) ? \\
$1P_1$ & 5.831 & 5.86 & 5.849 &         \\
$1P_1$ & 5.859 & 5.86 &       &         \\
$1P_0$ & 5.841 & 5.83 &       &          \\
$2S_0$ & 5.971 & 6.27 &       &          \\
$2S_1$ & 5.984 & 6.34 &       &          \\
\hline
\end{tabular}
\end{table}

Let us compare the obtained results with model independent
predictions of heavy quark effective theory (HQET).
Heavy quark symmetry provides relations between excited states
of $B$ and $D$ mesons, such as
\begin{equation}
\label{rel1}
\bar M_{B_1}-\bar M_{D_1}=\bar M_{B_{s1}}-\bar M_{D_{s1}}=
\bar M_{B}-\bar M_{D}=\bar M_{B_s}-\bar M_{D_s}=m_b-m_c,
\end{equation}
where $\bar M_{B_1}=(3M_{B_1}+5M_{B_2})/8$, $\bar M_{B}=(M_B+3M_{B^*})/4$ 
are appropriate spin averaged $P$- and $S$-wave states. 
We get from Tables~\ref{md}-\ref{mbs} the following
values of mass splittings
\begin{eqnarray}
&&\bar M_{B_1}-\bar M_{D_1}=3.29\ {\rm GeV}\qquad
\bar M_{B_{s1}}-\bar M_{D_{s1}}=3.30\ {\rm GeV}\cr
&&\bar M_{B}-\bar M_{D}=3.34\ {\rm GeV}\qquad 
\bar M_{B_s}-\bar M_{D_s}=3.33\ {\rm GeV},
\end{eqnarray}
in agreement with (\ref{rel1}). There arise also the following
relations between hyperfine splittings of levels
\begin{equation}
\Delta M_{B}\equiv M_{B_2}-M_{B_1}=\frac{m_c}{m_b}\Delta
M_{D}\equiv \frac{m_c}{m_b}\left(M_{D_2}-M_{D_1}\right),
\end{equation}
and the same for $B_s$ and $D_s$ mesons as well as for $P_1-P_0$ states.
Our model predictions for these splittings are displayed in 
Table~\ref{splt}.
\begin{table}[htb]
\caption{Hyperfine splittings of $P$ levels. All values are given in MeV.}
\label{splt}
\begin{tabular}{cccc|ccc}
\hline
States & $\Delta M_D$ & $\frac{m_c}{m_b}\Delta M_D$ & $\Delta M_B$ &
$\Delta M_{D_s}$ & $\frac{m_c}{m_b}\Delta M_{D_s}$ & $\Delta M_{B_s}$\\
\hline
$1P_2-1P_1$ & 45 & 14 & 14 & 45 & 14 & 13\\
$1P_1-1P_0$ & 63 & 20 & 19 & 61 & 19 & 18\\
\hline
\end{tabular}
\end{table}

In Tables~\ref{md}-\ref{mbs} we compare our relativistic quark
model results for heavy-light meson masses with the 
predictions of other quark  models of Godfrey
and Isgur \cite{gi}, Isgur \cite{i},  Eichten, Hill and Quigg 
\cite{ehq} and experimental data \cite{PDG,orb}. 
All these quark models  use the
expansion in inverse powers both of the heavy $m_Q$ and light $m_q$
quark masses for the $Q\bar q$ interaction potential. In ref.~\cite{gi} 
some relativization of the potential has been put in by hand, such
as relativistic smearing of coordinates and replacing the factors
$1/m_q$ by $1/{\epsilon_q(p)}$. However, the resulting potential in
this approach accounts only for some of the relativistic effects, 
while the others, which are of the same order of magnitude, are missing.
The considerations of Refs.~\cite{i,ehq} are closely related. The 
heavy quark expansion is extended to light ($u,d,s$) quarks and the
experimental data on $P$ wave masses of $K$ mesons are used to obtain
predictions for $B$ and $D$ mesons.

In the paper~\cite{i} it is argued that the heavy quark spin $P$-wave
multiplets
with $j=1/2$ ($0^+,1^+$) and $j=3/2$ ($1^+,2^+$) in $B$ and $D$ mesons
are inverted. 
The $2^+$ and $1^+$ states lie about 150 MeV below the 
$1^+$ and $0^+$ states.  In the limit $m_Q\to \infty$, we find 
the same inversion of these multiplets in our model, but the gap between
$j=1/2$ and $j=3/2$ states is smaller ($\sim 90$ MeV for $B$ and $D$ mesons
and $\sim 70$ MeV for $B_s$ and $D_s$ mesons), and  $1/m_Q$ corrections
reduce this gap further. However, the hyperfine splittings among the 
states in these multiplets turn out to be larger than in \cite{i}. 
As a result, the states from
the multiplets for $D$, $D_s$ and $B_s$ mesons
overlap in our model, however the heavy quark spin
averaged centres are still inverted. 
We obtain the following ordering of
$P$ states (with masses increasing from left to right): 
$B$ meson --- 
$1P_1(\frac32)$, $1P_2$, $1P_0$, $1P_1(\frac12)$;
$D_s$ meson --- $1P_0$, $1P_1(\frac32)$, $1P_2$, 
$1P_1(\frac12)$; $D$ and $B_s$
mesons --- $1P_1(\frac32)$, $1P_0$, $1P_2$, $1P_1(\frac12)$. Thus only for
$B$ meson we get the purely inverted pattern. Note that the model
\cite{gi} predicts the ordinary ordering of levels.
The results of our model agree well with available experimental data. 

{\bf 3. Semileptonic $B$ decays to orbitally excited $D$ mesons.}
In ref.~\cite{smexc} we have applied the relativistic quark 
model to the consideration 
of semileptonic decays of $B$ mesons to orbitally 
excited charmed mesons in the
leading order of the heavy quark expansion. At the leading order 
of the heavy quark expansion the Isgur-Wise \cite{iw1} 
functions $\tau_{3/2}$ and
$\tau_{1/2}$ are as follows
\begin{eqnarray}
\label{tau3}
\tau_{3/2}(w)&=&\frac{\sqrt{2}}{3}\frac{1}{(w+1)^{3/2}}\int\frac{d^3 p}
{(2\pi)^3}\bar\psi_{D(3/2)}({\bf p}+\frac{2\epsilon_q}{M_{D(3/2)}(w+1)}
{\bf \Delta})\cr
&&\times\left[-2\epsilon_q
\overleftarrow{\frac{\partial}{\partial p}}+\frac{p}{\epsilon_q+m_q}\right]
\psi_B({\bf p}),\\
\label{tau1}
\tau_{1/2}(w)&=&\frac{1}{3\sqrt{2}}\frac{1}{(w+1)^{1/2}}\int\frac{d^3 p}
{(2\pi)^3}\bar\psi_{D(1/2)}({\bf p}+\frac{2\epsilon_q}{M_{D(1/2)}(w+1)}
{\bf \Delta})\cr
&&\times\left[-2\epsilon_q
\overleftarrow{\frac{\partial}{\partial p}}-\frac{2p}{\epsilon_q+m_q}\right]
\psi_B({\bf p}),
\end{eqnarray}
where the arrow over $\partial/\partial p$ indicates that the derivative
acts on the wave function of the $D^{**}$ meson. The last  terms in 
the square brackets of these
expressions result from the wave function transformation  
associated with the relativistic  rotation of the
light quark spin (Wigner rotation) in
passing to the moving reference frame. These terms are numerically important
and lead to the suppression of the $\tau_{1/2}$ form factor compared to 
$\tau_{3/2}$. Note that if we  
had applied a simplified nonrelativistic quark model \cite{iw1}
these important contributions would be missing. Neglecting further the
small difference between the wave functions $\psi_{D(1/2)}$ and 
$\psi_{D(3/2)}$, the following relation between $\tau_{3/2}$ and 
$\tau_{1/2}$ would have been obtained \cite{llsw}
\begin{equation}\label{taunr}
\tau_{1/2}(w)=\frac{w+1}{2}\tau_{3/2}(w). 
\end{equation}
At the point $w=1$, where the initial $B$ meson and final $D^{**}$ are
at rest, we find instead the relation
\begin{equation}\label{diftau}
\tau_{3/2}(1)-\tau_{1/2}(1)\cong \frac12\int\frac{d^3p}{(2\pi)^3}
\bar\psi_{D^{**}}({\bf p})\frac{p}{\epsilon_q+m_q}\psi_B({\bf p}),
\end{equation}
obtained by assuming $\psi_{D(3/2)}\cong\psi_{D(1/2)}\cong\psi_{D^{**}}$.

In Table~\ref{tauv} we present our numerical results for $\tau_{j}(1)$ 
and its slope $\rho_j^2$ in comparison with other model predictions 
\cite{llsw,mlopr,ddgnp,w,cdp,gi,cccn}. We see that most of 
the above approaches predict close values for the function $\tau_{3/2}(1)$ 
and its slope $\rho_{3/2}^2$, while the results for $\tau_{1/2}(1)$ 
significantly differ from each other. This difference is a consequence
of a different treatment of the relativistic quark dynamics. Nonrelativistic
approaches predict $\tau_{3/2}(1)\simeq\tau_{1/2}(1)$ (see (\ref{taunr})),
while the relativistic treatment leads to $\tau_{3/2}(1)>\tau_{1/2}(1)$
(see (\ref{diftau})). Our results  for the branching ratios of 
$B\to D_{1,2}(3/2)e\nu$ decays are consistent with available experimental
data.
\begin{table}
\caption{The comparison of our model results for
the values of the functions $\tau_j$ at zero 
recoil of final $D^{**}$ meson  and their slopes $\rho_j^2$ with other
predictions.} 
\label{tauv}
\begin{tabular}{cccccccc}
\hline
   & our & \cite{llsw} & \cite{ddgnp} & \cite{w} & \cite{cdp}& 
\cite{mlopr},\cite{gi} & \cite{mlopr},\cite{cccn}\\
\hline
$\tau_{3/2}(1)$ & 0.49 & 0.41 & 0.56 & 0.66 &     & 0.54 & 0.52\\
$\rho_{3/2}^2$  & 1.53 & 1.5  & 2.3  & 1.9  &     & 1.5  & 1.45\\
$\tau_{1/2}(1)$ & 0.28 & 0.41 & 0.09 & 0.41 &$0.35\pm0.08$ & 0.22 & 0.06\\
$\rho_{1/2}^2$  & 1.04 & 1.0  & 1.1  & 1.4  &$2.5\pm1.0$ & 0.83 & 0.73\\ 
\hline
\end{tabular}
\end{table}

The work was supported  by the Deutsche Forschungsgemeinschaft
under contract Eb~139/1-3 and by the Russian Foundation for Fundamental
Research under Grant No.~96-02-17171.

\end{document}